\documentclass[prl,twocolumn,superscriptaddress]{revtex4}
\usepackage{amssymb}

\usepackage{amsmath,amsfonts,amssymb,epsfig,color}

\newcommand{\SO}[1]{\ensuremath{\mathrm{SO}( #1 )}}

\newcommand{\Spn}[1]{\ensuremath{\mathrm{Sp}( #1 )}}

\newcommand{\SUq}[1]{\ensuremath{\mathrm{SU}_q( #1 )}}

\newcommand{\spn}[1]{\ensuremath{\mathfrak{sp}( #1 )}}

\newcommand{\spq}[1]{\ensuremath{\mathfrak{sp}_q( #1 )}}

\newcommand{\suq}[2]{\ensuremath{\mathfrak{su}_q^{ #1 }( #2 )}}

\newcommand{\half}{\ensuremath{\textstyle{\frac{1}{2}}}}

\newcommand{\fpg}{\ensuremath{1f_{5/2}2p_{1/2}2p_{3/2}1g_{9/2}} }

\begin{document}

\title{Physical Significance of $q$-Deformation and Many-body
Interactions in Nuclei}
\author{K.D.~Sviratcheva}
\affiliation{Department of Physics and Astronomy, Louisiana State
University, Baton Rouge, LA 70803, USA}
\author{C.~Bahri}
\affiliation{Department of Physics and Astronomy, Louisiana State
University, Baton Rouge, LA 70803, USA}
\affiliation{Departemen Fisika, FMIPA, Universitas Indonesia, Depok 16424, Indonesia}
\author{A.I.~Georgieva}
\affiliation{Department of Physics and Astronomy, Louisiana State
University, Baton Rouge, LA 70803, USA}
\affiliation{Institute of Nuclear Research and Nuclear Energy,
Bulgarian Academy of Sciences, Sofia 1784, Bulgaria}
\author{J.P.~Draayer}
\affiliation{Department of Physics and Astronomy, Louisiana State
University, Baton Rouge, LA 70803, USA}


\begin{abstract}
The quantum deformation concept is applied to a study of pairing
correlations in nuclei with mass $40\leq A\leq 100$.  While the
nondeformed
limit of the theory provides a reasonable
overall description of certain nuclear properties and fine structure effects,
the results show that the $q$-deformation plays a
significant role in understanding higher-order effects in the many-body
interaction.
\end{abstract}

\maketitle


\noindent
\textit{1.~Introduction} ---
First used in applications of quantum inverse scattering \cite{KulishR83},
quantum (or $q$-) deformation \cite{Jimbo,Drinfeld} has been the focus of
considerable attention in various fields of physics in recent years. In
addition to purely mathematical examinations, recent studies of
interest include applications in string/brane theory, conformal field
theory, statistical/quantum mechanics, and metal clusters
\cite{BallesterosBH03,BaisSS02,Alg02,Zhang00,Bon00}, as well
as in nuclear physics \cite{Bon02,Ballesteros0203}.

Mathematically, a deformation parameter ($q$) is used to realize a
mapping of $c$-numbers (or operators) $X$ into their $q$-equivalents:
$[X]_{p} \doteq \allowbreak \frac{q^{p X}-q^{-p X}}{ q^{p}-q^{-p}} \allowbreak
\stackrel{q\rightarrow 1}{\rightarrow }X$ (denoted $[X]$ when $p=1$).
$[X]_{p}$ is nonlinear in $X$. It is invariant under the 
transformation $q
\rightarrow q^{-1}$, i.e. with respect to the sign of a real 
parameter $\varkappa $
(where $q=e^\varkappa$ for $q$ real), and hence depends only on even powers of
$\varkappa $.  A feature of any quantum algebra is that in the 
$\varkappa \rightarrow
0$  ($q \rightarrow 1$) limit, one recovers the nondeformed results,
just as classical mechanics (Galilean relativity) is restored from 
quantum mechanics (Einsteinian relativity)
when $\hbar \rightarrow 0$ ($\frac{1}{c}\rightarrow 0 $).

The earliest applications of the quantum algebraic concept to nuclear structure
were related to an \SUq{2} description of rotational bands in axially deformed
nuclei \cite{RRS90} and of like-particle pairing \cite{SharmaBon92}.  Even
though optimum values of 
the  $q$-parameter 
have achieved an overall improved
fit to the experimental energies, the question on the physical nature of
$q$-deformation when applied to the nuclear many-body problem remains open.

It is well-known that effective two-body interactions in nuclei are dominated
by pairing and quadrupole terms.  The former accounts for pair formation and
gives rise to a pairing gap in nuclear spectra, and the latter is responsible
for enhanced electric quadrupole transitions in collective rotational bands.
Indeed, within the framework of the harmonic oscillator shell-model,
both limits have a clear algebraic structure in the sense that the spectra
exhibit a dynamical symmetry. In the pairing limit the ``quasi-spin"
symplectic 
\Spn{4} ($\sim \SO{5}$)
\cite{Ker,Helmers,Hecht,Ginocchio} together with its dual
\Spn{2\Omega},
for $2\Omega$ shell degeneracy, use the seniority quantum number \cite{Rac,Flowers} to
classify the spectra.  On the other
hand, in the quadrupole limit symplectic \Spn{6,\mathbb{R}}
\cite{Ros} governs a shape-determined dynamics.

Pairing, introduced in physics for describing superconductivity, is 
fundamental to condensed matter, nuclear, and astrophysical phenomena of recent interest.
In nuclear physics, a two-body microscopic model with
\Spn{4} dynamical symmetry allows one to focus on like-particle 
($pp$ and $nn$) 
and proton-neutron $pn$ isovector
(isospin $T=1$) pairing correlations
and, in addition, to include a $pn$ isoscalar
($T=0$) interaction. Nuclear properties are generally well-described 
within this
framework
\cite{Svi02,SGD03stg}, with labels of the
\Spn{4} scheme yielding a basic understanding of the overall systematics.

The new feature reported on in this article is an extension of the
theory to include nonlinear local deviations from the pairing 
solution as realized
through $q$-deformation of the \spn{4}
algebra \footnote{We use lowercase (capital) letters for algebras (groups).}.
An important property of the $q$-deformed model is that it
does not violate physical laws fundamental to a quantum mechanical nuclear
system and conserves the angular momentum, the total number of
particles, and the isospin projection.

The quantum extension of \spn{4} makes possible the analytical 
modeling of a set
of many-body interactions. In general, the latter are rather complicated
to handle, nevertheless, they introduce an overall
improvement of the theory
\cite{Navratil9903Vesselin03}.
We aim to show that the $q \ne 1$
results are uniformly superior to those of the nondeformed limit and that the
$q$-parameter varies smoothly with nuclear characteristics. The 
results of this study
suggest that $q$-deformation has physical significance over-and-above 
the simple
pairing gap concept, extending to the very nature of the nuclear 
interaction itself.
The role of
$q$-deformation is not model limited, it can extend to include a description
of various many-body effects.

\vspace{0.2cm}
\noindent\textit{2.~Nonlinear pairing model} ---
The $\spq{4}$ deformed algebra \cite{Hayashi90,Sel95,fermRealSp4} is
realized in terms of $q$-deformed fermion
operators,
$\alpha^\dagger_{\nu =\{jm\sigma \}}$ and $\alpha_{\nu }$,
each of which creates and annihilates
a nucleon with isospin $\sigma$ ($\pm \half$ for proton/neutron) in a
single-particle state of total angular momentum $j$ (half-integer) 
with third projection
$m$.
The $q$-operators are defined through their anticommutation relations,
$
\{ \alpha_{jm\sigma}, \alpha^\dagger_{kn\tau} \}_{q^{\pm \delta _{\sigma \tau}}} =
     q^{\pm \frac{N_{2\sigma}}{2\Omega}} 
\delta _{jk}\delta _{mn} \allowbreak\delta _{\sigma \tau},
\{ \alpha^{(\dagger )}_{\nu }, \alpha^{(\dagger )}_{\nu ^\prime} \} = 0$,
where the $q$-anticommutator is $\{A,B\}_{q^p}= AB + q^p BA$,
$N_{2\sigma =\pm 1}$ is the proton (neutron) number operator and 
$2\Omega = \sum_j
(2j+1)$ is the space dimension for given $\sigma$
\cite{fermRealSp4}.
The  nondeformed $c^{(\dagger )}_{\nu }$
operators ($\alpha^{(\dagger )}_{\nu } \stackrel{q\rightarrow 1}{\rightarrow }
c^{(\dagger )}_{\nu }$) obey the usual anticommutation relations.
The basis operators, $T_{\pm}$ and $A^{(\dagger)}_{1,0,-1}$, of
the \spq{4} algebra are constructed as eight bilinear products of the
fermion $q$-operators coupled to total angular momentum and parity
$J^\pi=0^+$,
in addition to the nucleon number
operators $N_{\pm 1}$, which remain undeformed \cite{fermRealSp4}. The
isospin projection operator $T_0=\half(N_{+1} - N_{-1})$ and the 
total nucleon number
operator $N=N_{+1} + N_{-1}$ are also undeformed. In the $q \rightarrow 1$ limit,
$T_{0,\pm}$ are associated with isospin and $A^{(\dagger)}_{1,0,-1}$
create (annihilate) a proton-proton, proton-neutron, or 
neutron-neutron  $J=0$ pair.

As for the microscopic nondeformed approach,
the most general Hamiltonian \cite{Svi02} of a system with 
$q$-deformed symplectic
dynamical symmetry ($\spq{4} \supset \suq{}{2}$) and conserved proton 
and neutron particle numbers can be expressed as
\begin{widetext}
\begin{equation}
H_q =-\varepsilon _q N - G_q \sum _{k=-1}^1A^\dagger_k A_k
- F_q A^\dagger_0 A_0
- \textstyle{
\frac{E_q}{2\Omega } \left(\mathbf{T}^2-\Omega \left[ \frac{N}{2\Omega}\right]
\right)
-D_q \Omega \left[\frac{1}{\Omega }\right] \left[ T_0 \right]^2 
_{\frac{1}{2\Omega }}
-C_q 2  \Omega \left[\frac{1}{\Omega }\right]
{[\frac{N}{2} ] _{\frac{1}{2\Omega }} [\frac{N}{2}-2\Omega ] 
_{\frac{1}{2\Omega }} }},
\label{qH}
\end{equation}
\end{widetext}
where  $\mathbf{T}^2=
\Omega (\{T_+,T_-\}+ \left[\frac{1}{\Omega 
}\right][T_0]_{\frac{1}{2\Omega }}^{2})$
and the definitions of $[X]_p$ and $[X]$ are used.
In principle, the 
deformation parameters
$\gamma _q=\{ \varepsilon _q, G _q, F _q, E _q, D _q, C _q \}$ can  differ from
their nondeformed counterparts $\gamma =\{ \varepsilon , G, F, E, D, C \}$,
which we assume to be constant for all nuclei within a major shell.
The model describes the behavior of $N_{+1}$ valence protons and 
$N_{-1}$ valence
neutrons in the mean-field of a doubly-magic nuclear core. The
basis states, specified by the numbers of
$pn$ and like-particle pairs, are constructed by the
action of $A^\dagger _{0,\pm 1}$ on the vacuum.

The nondeformed Hamiltonian $H$, $H_q \stackrel{q\rightarrow 1}{\rightarrow } H$, is an
effective two-body interaction that includes isovector pairing (parameter $G$) and a
so-called symmetry term ($E$), which together with the
$N^2$-term arise naturally from a general two-body rotational and isospin
invariant microscopic interaction. Both $C$- and $E$-terms 
account for an isoscalar $pn$ interaction
that is diagonal in an isospin basis \footnote{In addition,
the $F$-term accounts for a possible, even if extremely small, 
isospin mixing.}.
These interactions govern the lowest $0^+$ isobaric analog states of light and
medium mass even-$A$ nuclei
($40\le A\le 100$) with protons and neutrons occupying the same major
shell, where the seniority zero limit is approximately valid
\cite{Svi02,SGD03stg,EngelLV96}. For these states,
the nondeformed model has already proven to provide a
reasonable overall description for a total of 136 nuclei \cite{Svi02}. This
includes a remarkable reproduction of the energy of the states and 
their detailed
structure reflecting observed
$N_{+1}=N_{-1}$ irregularities and staggering patterns
\cite{SGD03stg}.  As a consequence, any deviation within a nucleus from the
reference global behavior can be attributed to local effects
which although typically small
can be important for determining the detailed structure of individual 
nuclei and
hence need to be taken into account \cite{Navratil9903Vesselin03}.

As a group theoretical approach, the quantum extension of $H$ 
includes many-body
interactions in a very 
prescribed way,
retaining the simplicity of the exact solution.
Moreover, the quantum model not only has the
$\spq{4} \supset \suq{}{2}$ dynamical symmetry,
it contains the original dynamical \Spn{4} symmetry.

\noindent\textit{3.~Novel properties of $q$-deformation} --- From an 
undeformed perspective, the deformation introduces higher-order, many-body terms 
into a theory
that starts with only one-body and two-body interactions. The way in which the
higher-order effects enter into the theory is governed by the $[X]$ 
form. In terms of
$\varkappa$, 
everything is
tied to the deformation with
$\textstyle{[X] =\frac{\sinh{(\varkappa  X )}}{\sinh{(\varkappa )}}=
X(1+\varkappa ^{2} \frac{X^{2}-1}{6} +}\allowbreak
\textstyle{\varkappa ^{4} \frac{3X^{4}-10X^{2}+7}{360}
+...)  \stackrel{\varkappa \rightarrow 0}{\rightarrow }X}$. An
illustrative example is the expansion in $\varkappa $ of the last term in $H_q$
(\ref{qH}),
$-C_q 2  \Omega \left[\frac{1}{\Omega }\right]
[\frac{N}{2} ] _{\frac{1}{2\Omega }}
[\frac{N}{2}-2\Omega ] _{\frac{1}{2\Omega }}
=
-2C_q \frac{N}{2}(\frac{N}{2}-2\Omega ) -\frac{C_q \varkappa ^2
\{(16\Omega ^2-24\Omega +5)(V^{(1)}+V^{(2)})
+6V^{(2)}
+(6-8\Omega )V^{(3)}
+ V^{(4)}
\}} {96\Omega^2}-...$, with $V^{(1)}=
\sum _{\nu_1} c^\dagger _{\nu_1} c_{\nu_1} $,
$V^{(2)}=
\sum _{\nu_1 \nu _2}c^\dagger _{\nu_1} c^\dagger _{\nu_2}c_{\nu_2}c_{\nu_1}$,
$V^{(3)}= \allowbreak 
\sum _{\nu_1 \nu _2 \nu_3 }
c^\dagger _{\nu_1} c^\dagger _{\nu_2} c^\dagger _{\nu_3}
c_{\nu_3} c_{\nu_2} c_{\nu_1} $,
and  $V^{(4)}=
\sum _{\nu_1 \nu _2 \nu_3 \nu_4 } \allowbreak
c^\dagger _{\nu_1} c^\dagger _{\nu_2} c^\dagger _{\nu_3} c^\dagger _{\nu_4}
c_{\nu_4} c_{\nu_3} c_{\nu_2} c_{\nu_1}$. The zeroth-order
approximation corresponds to the nondeformed two-body force 
and coincides
with it for a strength $C_q$ equal to $C$, and the higher-order terms introduce
many-body interactions. The latter may not be negligible, for 
example, our results show
that the energy contribution of the four-body interaction in the expansion above
can reach a magnitude of several MeV in nuclei 
in the $\fpg $ shell.

Similarly, the zeroth-order term of $H_{q}$
(\ref{qH}) coincides with the $H$ nondeformed interaction
only if the strength parameters are equal, $\gamma _q=\gamma $. This term
must remain unchanged when deformation is introduced, since
$H$ has been shown to reproduce reasonably well the overall 
behavior common for
all the nuclei in a shell. This is why we fix the values of the parameters
$\gamma _q=\gamma $ and allow only $\varkappa$ to vary. The decoupling of the
deformation from the $\gamma $ parameters that are used to characterize the
two-body interaction itself, means that the latter can be assigned 
best-fit global
values for the model space under consideration without compromising 
overall quality
of the theory. This in turn underscores the fact that the deformation 
represents
something fundamentally different, a feature that cannot be  ``mocked up'' by
allowing the strengths of the nondeformed interaction to absorb its 
effect.  In
short, the $q$-deformation adds to the theory, which describes quite 
well the overall
nuclear behavior, a mean-field correction along with two-, three-, 
and many-body
interactions of a local character that can be responsible for 
residual single-particle
and many-body effects.

\begin{figure}[th]
\centerline{\hbox{\epsfig{figure=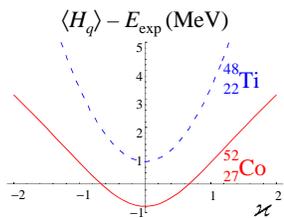,width=4cm}}}
\caption{Differences between theoretical and
experimental energies vs. the
$\varkappa$ parameter for a typical near-closed shell nucleus
(solid line) and for a mid-shell nucleus (dashed line). }
\label{fig:Hvskappa}
\end{figure}

\vspace{0.2cm}
\noindent\textit{4.~Analysis of the role of the $q$-deformation} ---
Since the $q$-parameter is associated with local phenomena, it is expected to
vary from one nucleus to another. The
possible presence of local effects
built over the global properties of the $0^+$ states under consideration can be
recognized within an individual
nucleus by the deviation of the predicted nondeformed energy $\langle
H\rangle $ from the experimental value $E_{\exp }$,
namely, the solution of the equation $\langle H_q\rangle
=E_{\exp }$ provides a rough estimate for $\varkappa $ (see
Fig.~\ref{fig:Hvskappa}).  However, in nuclei where
$\langle H\rangle  \geq E_{\exp }$ there is no solution (see
Fig.~\ref{fig:Hvskappa}) and the theoretical prediction closest to the
experiment occurs at the
nondeformed point, $\varkappa =0$.

\begin{figure}[th]
\centerline{\hbox{\epsfig{figure=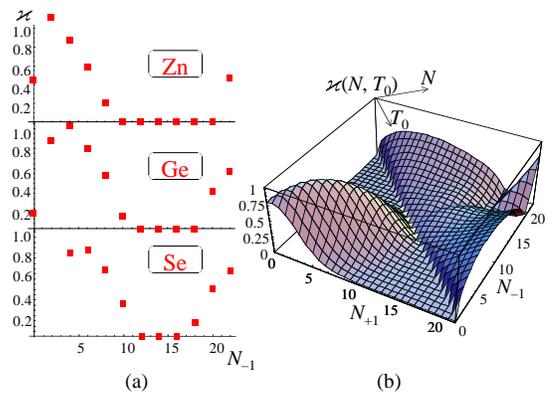,width=7.1cm}}}
\caption{$\varkappa $-Parameter estimation: (a) within
each nucleus,
and (b) $\varkappa
(N,T_0)$  within the \fpg shell (global parameters: $\varepsilon=13.851$,
$G/\Omega =0.296$,
$F/\Omega =0.056$,
$E/(2\Omega)=-0.489$, $D=-0.307$, and $C=0.190$ in MeV).}
\label{fig:KappaFnVsNmFit}
\end{figure}

The analysis yields values for the deformation parameter
$|\varkappa |$ for each nucleus that fall on a smooth curve (see
Fig.~\ref{fig:KappaFnVsNmFit}(a)). The observed smooth
behavior of $\varkappa $ 
reveals its functional dependence on the
model quantum numbers. This result, even though qualitative, 
underscores the fact
that the $q$-deformation as prescribed by the \spq{4} model is not random in
character but rather fundamentally related to the very nature of the nuclear
interaction.

This, in turn, allows us to assign a parametrized functional dependence of the
deformation parameter on the total particle number $N$ and the 
isospin projection $T_0$,
\begin{widetext}
\begin{equation}
\textstyle{ \varkappa (N,T_0)=\xi _1(\frac{N}{2\Omega  }-1)(\frac{N}{2\Omega
}+\xi _2-2\theta(N-2\Omega ))e^{-0.5(\frac{2T_0}{\xi _3})^2}
+\xi _4\theta(N-2\Omega )|T_0|\sqrt{\frac{N}{2\Omega }-1},} \quad
\theta(x)=\left\{
   \begin{array}{ll}
   1,\quad  x \geq 0 \\
   0,\quad  x<0
   \end{array}
\right. ,
\label{kappaFn}
\end{equation}
\end{widetext}
which reflects the complicated development of nonlinear effects
observed in Figure \ref{fig:KappaFnVsNmFit}(a).
As a next step, we use the $\varkappa (N,T_0) $ deformation function
(\ref{kappaFn}) to fit the minimum eigenvalues of
$H_q$ (\ref{qH}) to the relevant experimental energies of the 
even-even nuclei in
the  $1f_{7/2}$ and \fpg shells. In doing this, we
minimize any renormalization of the $q$-deformed parameter due
to a possible influence of other local effects that are not present in the
model. In the fitting procedure, only the four parameters 
($\xi_1,_2,_3,_4$) of $\varkappa (N,T_0)$ in Eq.(\ref{kappaFn}) are varied.
Determined statistically, they provide an estimate for the
overall significance of $q$-deformation within a 
shell.

\begin{figure}[th]
\centerline{\hbox{\epsfig{figure=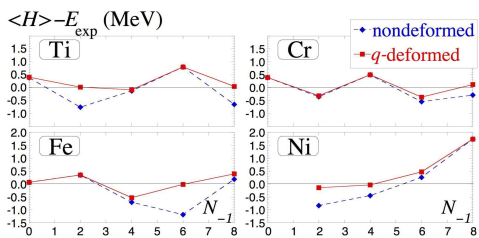,width=6.3cm}}}
\caption{The $q$-deformed and nondeformed
energies compared to experimental values for even-even isotopes in 
the $1f_{7/2}$ shell
(global parameters: $ \varepsilon=13.149$, $G/\Omega =0.453$,
$F/\Omega=0.072$, $E/(2\Omega)=-1.120$, $D=0.149$, and $C=0.473$
in MeV).}
\label{fig:BindEn}
\end{figure}

The $q \ne 1$ results are uniformly superior to those of the nondeformed
limit. In the \fpg shell, for example, the $q$-deformed model, $SOS_q =130.21$ MeV$^2$ 
($\chi _q=1.28$
MeV) \footnote{$SOS$ is defined as the sum of the squared differences 
in the theoretical
and experimental energies, and
$\chi ^2$ is the averaged $SOS$ per a degree of freedom in the
statistics. },
clearly improves the nondeformed theory, $SOS=271.63$ MeV$^2$ ($\chi=1.79$
MeV). The optimum results are achieved for:
\begin{equation}
\xi _1=-2.13,\quad \xi _2=0.37,\quad \xi _3=3.07,\quad \xi _4=0.15.
\label{kappaFnParamNi}
\end{equation}
The behavior of the $q$ deformation (as prescribed by Eq. (\ref{kappaFn})) is
consistent in both of the regions considered (shells $1f_{7/2} $ and 
$\fpg$), with a
general trend of higher values above  mid-shell, where the increase 
in particle number
can lead to stronger nonlinear effects. As a whole, the model with the local
$q$ improves the energy prediction compared to the nondeformed global model
and reproduces more closely the experiment (see Fig.~\ref{fig:BindEn}). One reason may be
that the $q$-deformed fermions, unlike usual quasiparticles,
indeed obey the fundamental laws.

The many-body nature of the interaction is most important away from 
mid-shell and
for many even-even nuclei tends to peak [with significant values of $q$] when
$N_{+1}=N_{-1}$ where strong pairing correlations are expected (see
Fig.~\ref{fig:KappaFnVsNmFit}).
Values of the deformation parameter $q\approx 1$ 
may be found
in nuclei with only one or two particle/hole pairs from a closed 
shell. For these
nuclei the number of particles is
insufficient to sample the effect of higher-order terms in a deformed 
interaction and
the nondeformed limit gives a good description.

Around mid-shell ($N\approx 2\Omega $) the deformation adds little 
improvement to the $\varkappa =0$ theory.
This suggests that for these nuclei the many-body interactions as prescribed
by $\varkappa (N,T_0)$ in Eq. (\ref{kappaFn}) are negligible
and the model is not sufficient to describe other types of local
effects that may be present. The results imply that even though the
$q$-parameter gives additional freedom for all the nuclei, it only improves
the model around regions of dominant pairing correlations.
In short, the pair formation favors the nonnegligible
higher-order interactions between the pair constituents that are detected via
the \spq{4} model.

A $q$-deformed nonlinear extension of the \Spn{4} model, which is 
the underlying
symmetry for describing
isovector
pairing correlations
and $pn$ isoscalar interactions in atomic nuclei, has been investigated.  When
compared to experimental data, the theory shows a smooth functional 
dependence of the
deformation parameter $q$ on the proton and neutron numbers. In 
addition, the $q
\ne 1$ results are uniformly superior to those of the nondeformed limit. 
The outcome suggests that the deformation has physical significance 
related to the very
nature of the nuclear interaction itself and beyond what can be 
achieved by simply
tweaking the  parameters of a two-body interaction.
The specific features of the nuclear structure can be investigated 
through the use
of a local $q$ that detects the presence and importance of
many-body interactions accompanying dominant pairing correlations in nuclei.
This is in
addition to the good description of the global properties of the nuclear
dynamics provided by the nondeformed two-body interaction.
Although the physical significance of $q$-deformation is presently
approached within the
\Spn{4} theoretical framework, it is clearly model independent and 
can reveal various
many-body phenomena.  The results also underscore the need for additional
studies to achieve a more comprehensive understanding of
$q$-deformation in  nuclear physics.

In summary, the concept of quantum  deformation
has been linked to the smooth behavior of physical phenomena in atomic nuclei.

This work was supported by the US National Science Foundation, Grant
Number 0140300.

\end{document}